\documentclass[manuscript]{aastex}

\shorttitle{On the Transition-Region lines observed by the IRIS}
\shortauthors{Dud\'{i}k et al.}

\begin{document}

\title{Solar Transition-Region Lines Observed by the Interface Region Imaging Spectrograph: \\ Diagnostics for the \ion{O}{4} and \ion{Si}{4} Lines}

\author{J. Dud\'{i}k\altaffilmark{1,2} and G. Del Zanna}
    \affil{DAMTP, CMS, University of Cambridge, Wilberforce Road, Cambridge CB3 0WA, United Kingdom}
    \email{J.Dudik@damtp.cam.ac.uk}
\author{E. Dzif\v{c}\'{a}kov\'{a}}
    \affil{Astronomical Institute of the Academy of Sciences of the Czech Republic, Fri\v{c}ova 298, 251 65 Ond\v{r}ejov, Czech Republic}
\author{H. E. Mason}
    \affil{DAMTP, CMS, University of Cambridge, Wilberforce Road, Cambridge CB3 0WA, United Kingdom}
\and
\author{L. Golub}
     \affil{Harvard-Smithsonian Center for Astrophysics, 60 Garden Street, Cambridge, Massachusetts, 01238, USA}

\altaffiltext{1}{DAPEM, Faculty of Mathematics Physics and Computer Science, Comenius University, Mlynsk\'{a} Dolina F2, 842 48 Bratislava, Slovakia}
\altaffiltext{2}{RS Newton International Fellow}

\begin{abstract}
The formation of the transition-region \ion{O}{4} and \ion{Si}{4} lines observable by the Interface Region Imaging Spectrograph (IRIS) is investigated for both Maxwellian and non-Maxellian conditions characterized by a $\kappa$-distribution exhibiting a high-energy tail. The \ion{Si}{4} lines are formed at lower temperatures than the \ion{O}{4} lines for all $\kappa$. In non-Maxwellian situations with lower $\kappa$, the contribution functions are shifted to lower temperatures. Combined with the slope of the differential emission measure, it is possible for the \ion{Si}{4} lines to be formed at very different regions of solar transition region than the \ion{O}{4} lines; possibly close to solar chromosphere. Such situations might be discernible by IRIS. It is found that photoexcitation can be important for the \ion{Si}{4} lines, but is negligible for the \ion{O}{4} lines. The usefulness of the \ion{O}{4} ratios for density diagnostics independently of $\kappa$ is investigated and it is found that the \ion{O}{4} 1404.78\AA\,/1399.77\AA~ratio provides a good density diagnostics except for very low $T$ combined with extreme non-Maxwellian situations.
\end{abstract}


%
%
%
\section{Introduction}
\label{Sect:1}

The solar transition-region (hereafter, TR), is the plasma with temperatures between chromospheric and coronal ones. While the transition-region emission can originate in closed magnetic loops that do not reach the corona, there are large regions of the strong magnetic field, such as plages, where the field lines can be ``open'' and reach into the corona. Such regions will be characterized by strong gradients of temperature and density. Under these conditions, the distribution of electron energies can depart from the Maxwellian one and become non-Maxwellian \citep[e.g.,][]{Roussel-Dupre80,Shoub83,Ljepojevic88}. Some indications of the presence of non-Maxwellian distributions were found from \ion{Si}{3} transition-region spectra \citep{Dufton84,Keenan89,Pinfield99}, while other authors \citep[e.g.,][]{Doschek97} did not find any.

\citet{Dzifcakova11Si} used the SUMER \ion{Si}{3} line intensities reported by \citet{Pinfield99} and performed diagnostics of temperature and density. It was found that different ratios do not yield consistent results if a Maxwellian distribution is assumed. Large discrepancies between theory and observations were found for active region spectra. The authors also performed the diagnostics under the assumption of $\kappa$-distributions \citep[e.g.,][]{Vasyliunas68,Owocki83,Livadiotis09}. Then, the consistency in terms of temperature and density was found only if the additional effect of photoexcitation by the photospheric radiation was included. The authors diagnosed $\kappa$\,=\,7 for the active region spectra, and $\kappa$\,=\,10$-$11 for the quiet-Sun, meaning that the active region TR is more non-Maxwellian than the quiet-Sun. Furthermore, these diagnostics were also shown to be valid for multithermal plasma characterized by a differential emission measure (DEM).

The Interface Region Imaging Spectrograph (IRIS) is a new, powerful instrument dedicated to the observations of the solar chromosphere and transition region. Its second far-ultraviolet channel (1390--1406\AA) contains several lines belonging to \ion{Si}{4}, \ion{O}{4} and \ion{S}{4} (Table \ref{Table:1}). We note that while direct diagnostics of the electron distribution function are not expected because of the similarity of the transitions observed and wavelength constraints of the instrument, the non-Maxwellian distributions can still have large effects on the formation of these lines. In this Letter, we investigate the formation of the transition region lines under non-Maxwellian conditions characterized by $\kappa$-distributions. We also investigate the effects of photoexcitation on line formation and the possible influence of non-Maxwellian effects on the density diagnostics known for Maxwellian distribution \citep[e.g.,][]{Doschek84}. Finally, we are not concerned in this paper with the formation of lines belonging to low ionization stages, such as \ion{O}{2}, \ion{C}{2} or others, since these may be heavily influenced by opacity effects in dense plasmas and no longer formed under optically thin or non-Maxwellian conditions. We also note that the response of the IRIS TR lines to non-equilibrium ionization already has been investigated by \citet{Doyle13} and \citet{Olluri13}.

%
%
\section{Method}
\label{Sect:2.1}
The $\kappa$-distributions are defined as a two-parametric distribution with parameters $T$ and $\kappa$ \citep[e.g.,][]{Owocki83,Livadiotis09}
\begin{equation}
	f_{\kappa}({\cal E})d{\cal E} = {\cal A}_{\kappa}~ \frac{2}{\pi^{1/2} (k_\mathrm{B}T)^{3/2}}~ \frac{{\cal E}^{1/2}d { \cal E} } { (1 + \frac { {\cal E}}{( \kappa - 1.5) kT})^{ \kappa + 1 }}\,,
	\label{Eq:kappa}
\end{equation}
where $k_\mathrm{B}$\,=\,1.38 $\times 10^{-16}$ ergs\,cm$^{-1}$ is the Boltzmann constant and $ {\cal A}_{\kappa} = \Gamma ( \kappa + 1 )/\left[ \Gamma (\kappa -0.5) ( \kappa - 1.5 )^{3/2}\right]$ is the normalization constant. We note that $\kappa$\,$\in$\,$\left(3/2,+\infty\right)$ and $T$ is defined in terms of mean energy $\left<{\cal E}\right>$ as $T$\,=\,$2\left<{\cal E}\right>/3k_\mathrm{B}$. Maxwellian distribution is recovered for $\kappa$\,$\to$\,$+\infty$. For finite $\kappa$, the distribution has increased number of both low-energy and high-energy electrons compared to Maxwellian distribution and exhibits a high-energy power-law tail and a near-Maxwellian core with temperature $T_\mathrm{core}$\,=\,$T (\kappa -3/2)/\kappa$ \citep{Oka13}. The differences with respect to Maxwellian increase with decreasing $\kappa$ \citep[see, e.g., Fig. 1 in][]{Dzifcakova13}.

\begin{figure*}
       \centering
       \includegraphics[width=8.0cm,clip]{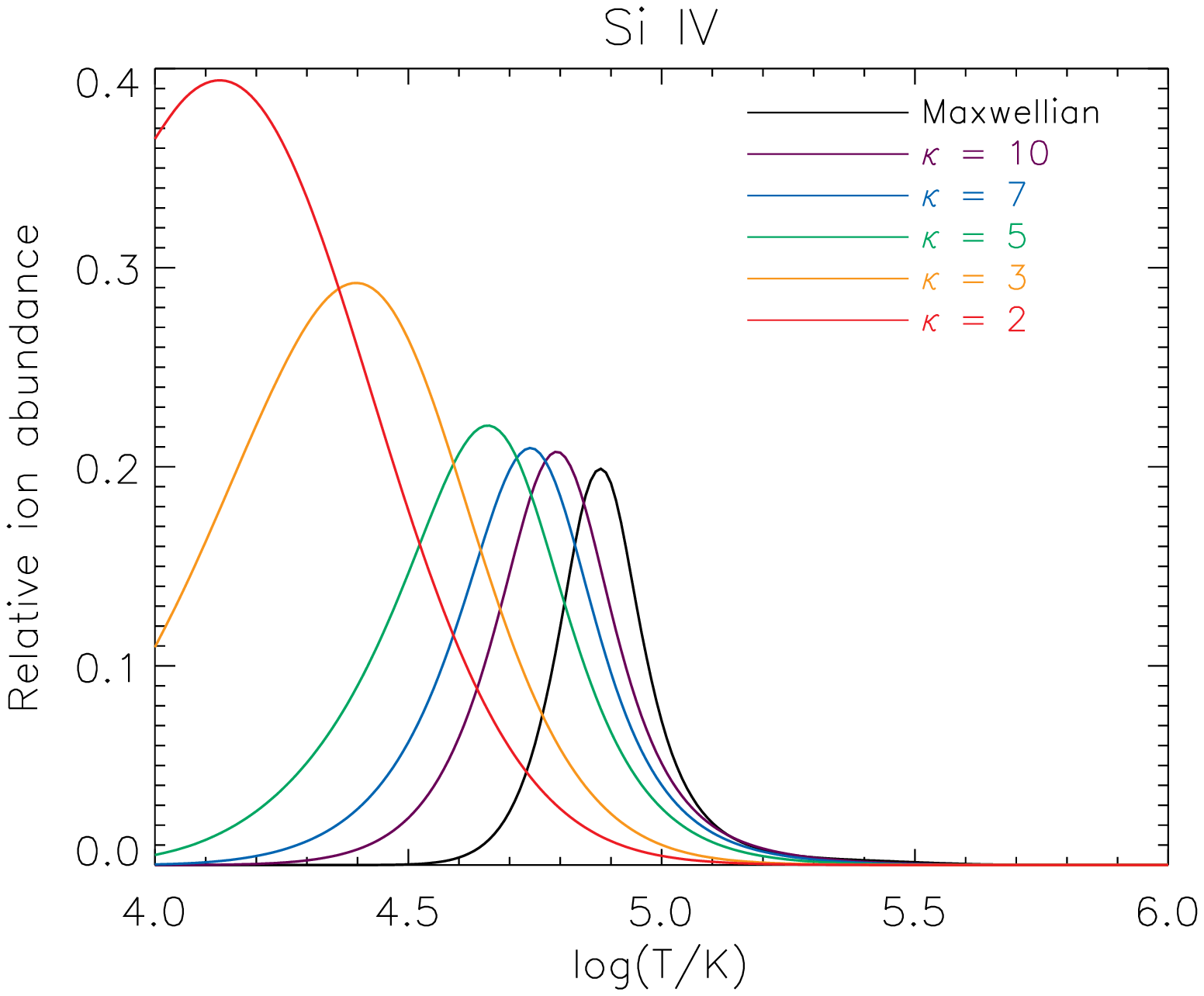}
       \includegraphics[width=8.0cm,clip]{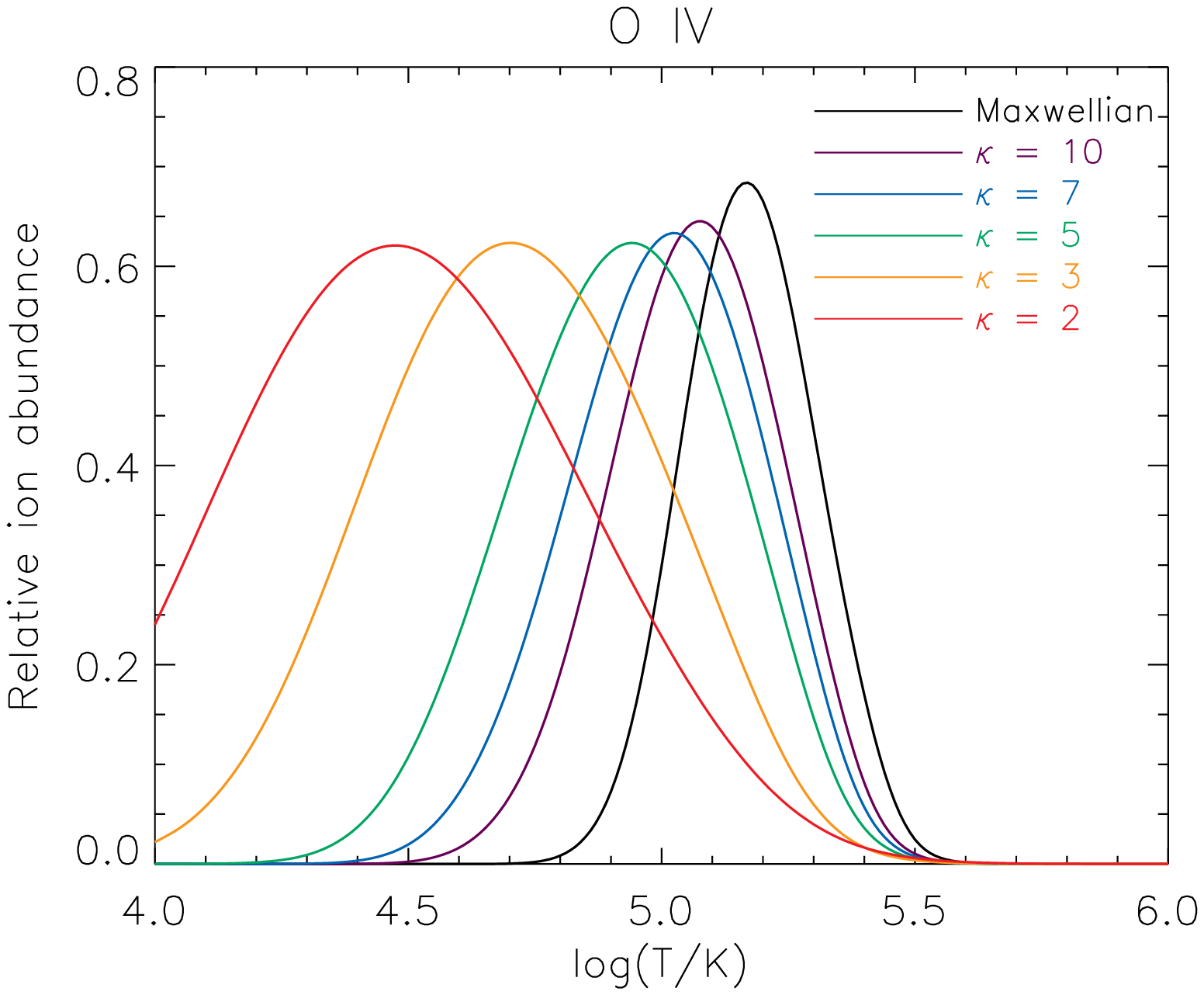}
       \includegraphics[width=8.0cm,clip]{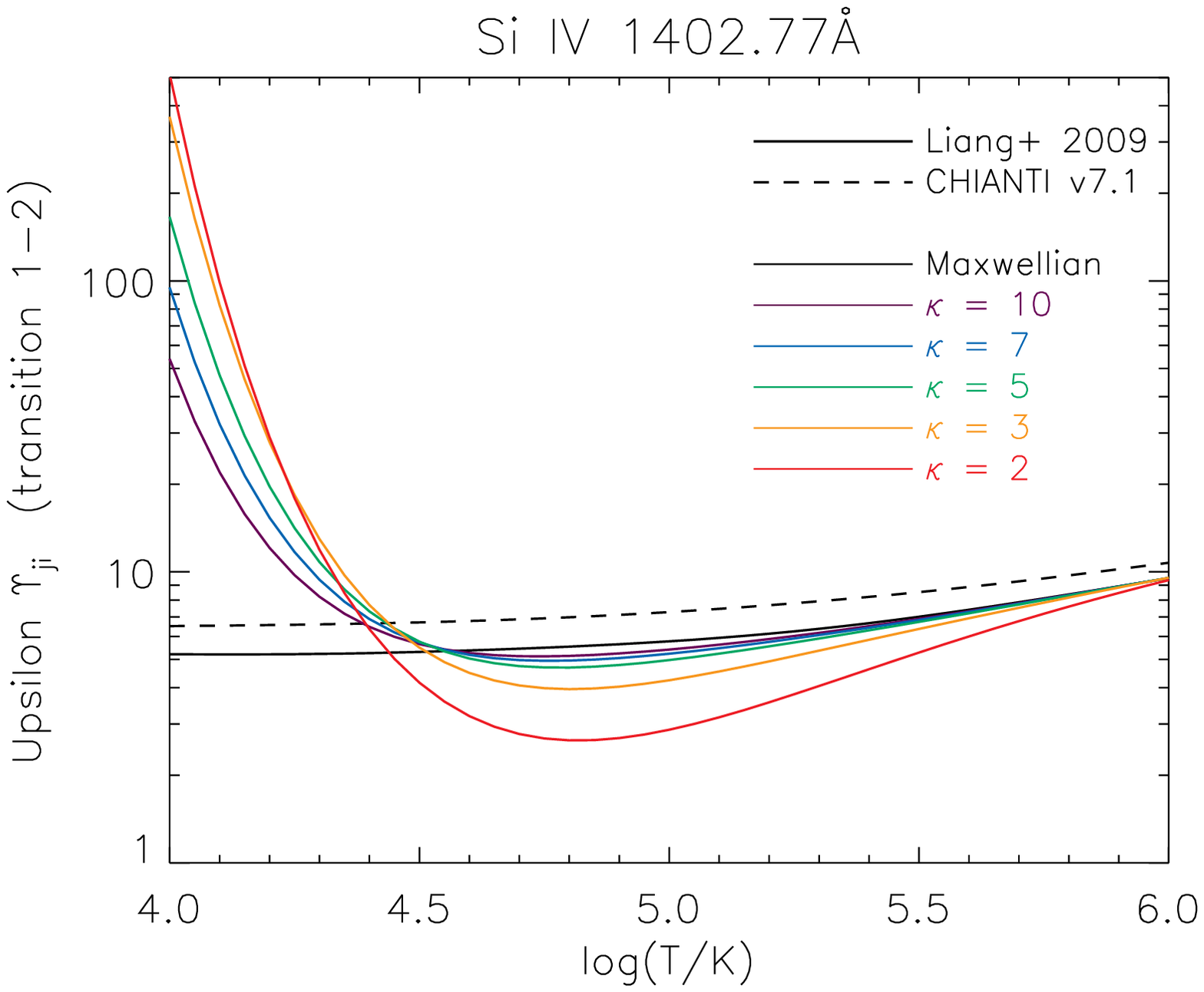}
       \includegraphics[width=8.0cm,clip]{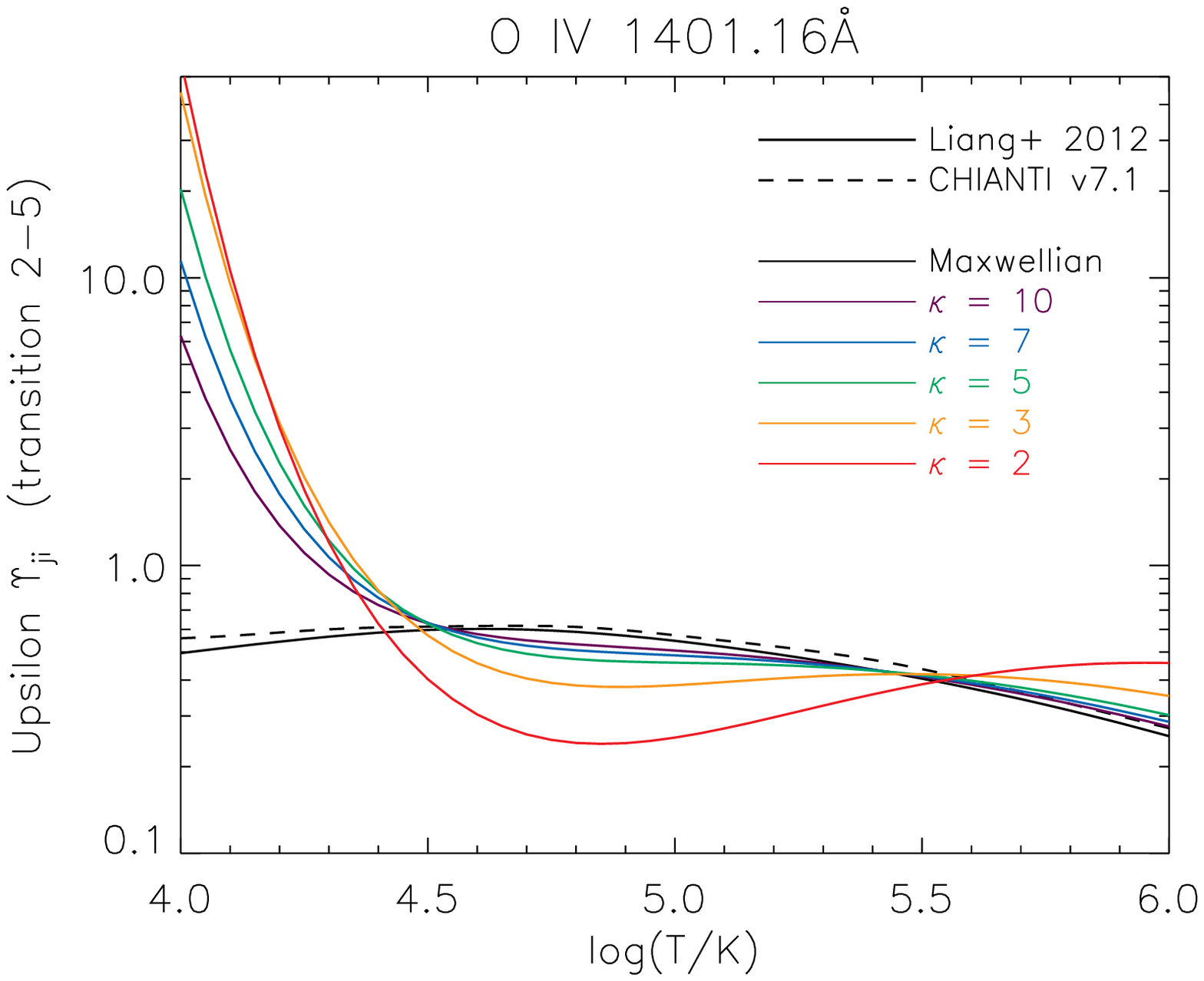}
     \caption{\textit{Top}: Relative ion abundances for \ion{Si}{4} (\textit{top left}) and \ion{O}{4} (\textit{top right}). \textit{Bottom}: Distribution-averaged collision strengths $\Upsilon_{ji}$ for the \ion{Si}{4} 1402.77\AA~and \ion{O}{4} 1401.16\AA~transitions. (A color version of this figure is available in the online journal.)}
       \label{Fig:atomic_data}
   \end{figure*}

The intensity $I_{ji}$ of an optically thin emission line with a wavelength $\lambda_{ji}$ can be written as \citep[c.f.,][]{Phillips08}
\begin{eqnarray}
	I_{ji} 	&=& \int G_{ji}(T,n_\mathrm{e},\kappa) A_X n_\mathrm{e}^2 \mathrm{d}l \\
		&=& \int G_{ji}(T,n_\mathrm{e},\kappa) A_X \mathrm{DEM}(T) \mathrm{d}T \,,
	\label{Eq:Int}
\end{eqnarray}
where $n_\mathrm{e}$ is the electron density, $A_X$ is the relative abundance, DEM$(T)$ is the differential emission measure along the line of sight $l$, defined as DEM$(T)$\,=\,$n_\mathrm{e}^2 \mathrm{d}l / \mathrm{d}T$. The $G_{ji}(T,n_\mathrm{e},\kappa)$ is the line contribution function, which can be expressed as
 \begin{equation}
	G_{ji}(T,n_\mathrm{e},\kappa) = \frac{hc}{\lambda_{ji}} \frac{A_{ji}}{n_\mathrm{e}} \frac{n_{X,i}^{+k}}{n_X^{+k}} \frac{n_X^{+k}}{n_X} \frac{n_\mathrm{H}}{n_\mathrm{e}}\,,
	\label{Eq:CF}
\end{equation}
where $hc/\lambda_{ji}$ is the photon energy, $A_{ji}$ is the Einstein coefficient for spontaneous emission, $n_\mathrm{H}$ is the hydrogen density, $n_X^{+k} / n_X$ is the relative abundance of the ion $+k$, and $n_{X,i}^{+k} / n_X^{+k}$ is the fraction of the ion $+k$ with the electron in the excited upper level $i$.
The $G_{ji}(T,n_\mathrm{e},\kappa)$ is a function of $\kappa$ because of the changes in both ionization equilibrium with $\kappa$ \citep{Dzifcakova02,Dzifcakova13} and excitation rates \citep[e.g.,][]{Dzifcakova06,DzifcakovaMason08}. The relative ion abundances, obtained from \citet{Dzifcakova13}, are shown in Fig. \ref{Fig:atomic_data}, \textit{top}.

We calculated the distribution-averaged collision strengths $\Upsilon_{ji}(T,\kappa)$ by integration of the collision strengths $\Omega_{ji}$ (non-dimensionalised excitation/de-excitation cross-sections) over the distribution function \citep{Seaton53}
\begin{equation}
	\Upsilon_{ji}(T,\kappa) = \frac{\sqrt\pi}{2} \mathrm{exp}\left({\frac{hc}{\lambda_{ji}k_\mathrm{B}T}}\right) \int_{0}^{\infty} \Omega_{ji}({\cal E}) \left(\frac{{\cal E}}{k_\mathrm{B}T}\right)^{-1/2} f_\kappa({\cal E}) \mathrm{d}{\cal E}\,.
	\label{Eq:Omega}
\end{equation}
The $\Omega_{ji}$ used here were calculated within the APAP network\footnote{www.apap-network.org} by \citet{Liang09} and \citet{Liang12} for \ion{Si}{4} and \ion{O}{4}, respectively. These represent the state-of-art atomic data and will be implemented in the upcoming CHIANTI v8. The numerical integration of the collision strengths for these ions was performed using a method similar to \citet{Bryans06}. We compared the $\Upsilon_{ji}(T,\kappa)$ obtained, shown in Fig. 1 \textit{bottom}, with the ones calculated by the approximative method of \citet{DzifcakovaMason08}. Excellent agreement within a few per cent was found. For comparison, previous atomic data calculations of \citet{Sampson90}, \citet{Martin95}, and \citet{Aggarwal08}, available within CHIANTI v7.1 \citep{Dere97,Landi13}, are also plotted in Fig.\,\ref{Fig:atomic_data}, \textit{bottom}. The differences for the \ion{O}{4} 1401.16\AA~line are up to 12\% and for the \ion{Si}{4} 1402.77\AA~line up to 26\%.

Our $G_{ji}(T,n_\mathrm{e},\kappa)$ for \ion{O}{4} and \ion{Si}{4} using $\Upsilon_{ji}(T,\kappa)$ from the original collision strengths are subsequently calculated using our own modification of the CHIANTI v7.1 software. For \ion{S}{4}, we utilize the atomic data of \citet{Kelleher99}, \citet{Tayal00}, and \citet{Hibbert02} available within CHIANTI v7.1 and the method of \citet{DzifcakovaMason08}, since no collision strength data are available. The abundances are taken from \citet{Asplund09}, i.e., are assumed to be photospheric. We also assume a contribution from photoexcitation by a Sun-like black-body, characterized by $T_\mathrm{eff}$\,=\,6000\,K and $R/R_\odot$\,=\,1.

%
%
%
\section{Results}
\label{Sect:3}
%
\subsection{Contribution functions for $\kappa$-distributions}
\label{Sect:3.1}

\begin{figure*}
       \centering
       \includegraphics[width=7.8cm,clip]{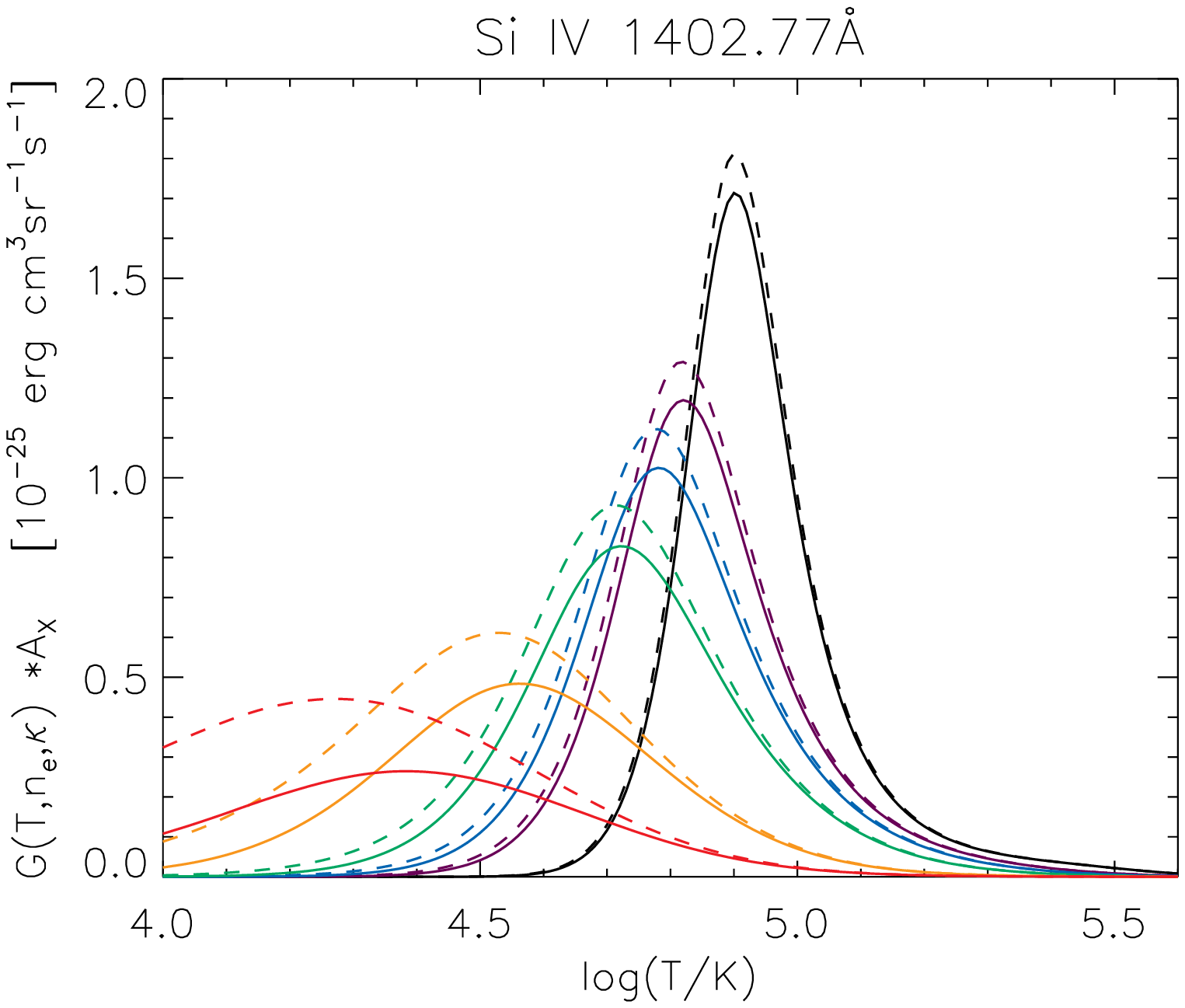}
       \includegraphics[width=7.8cm,clip]{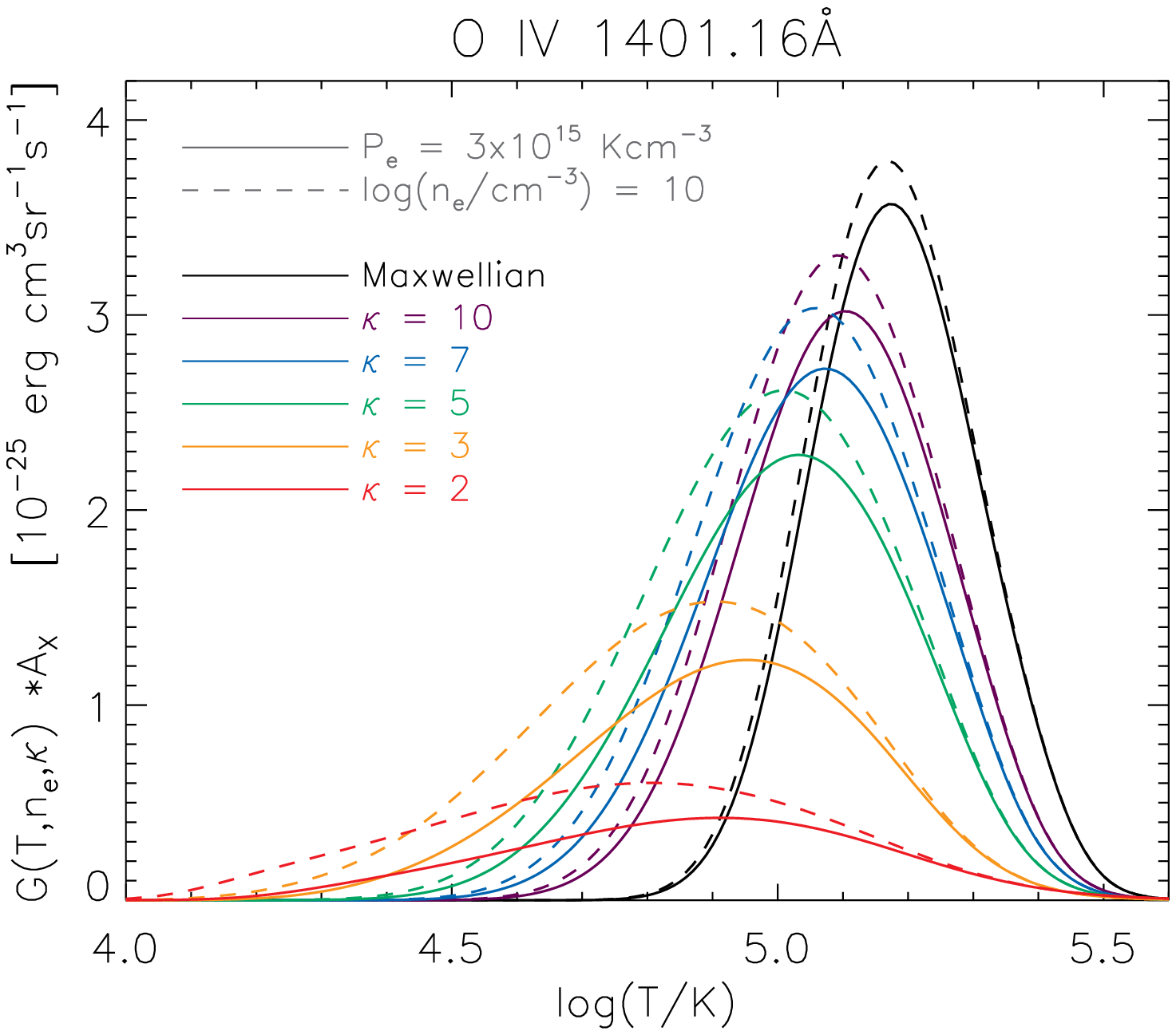}
       \includegraphics[width=7.8cm,clip]{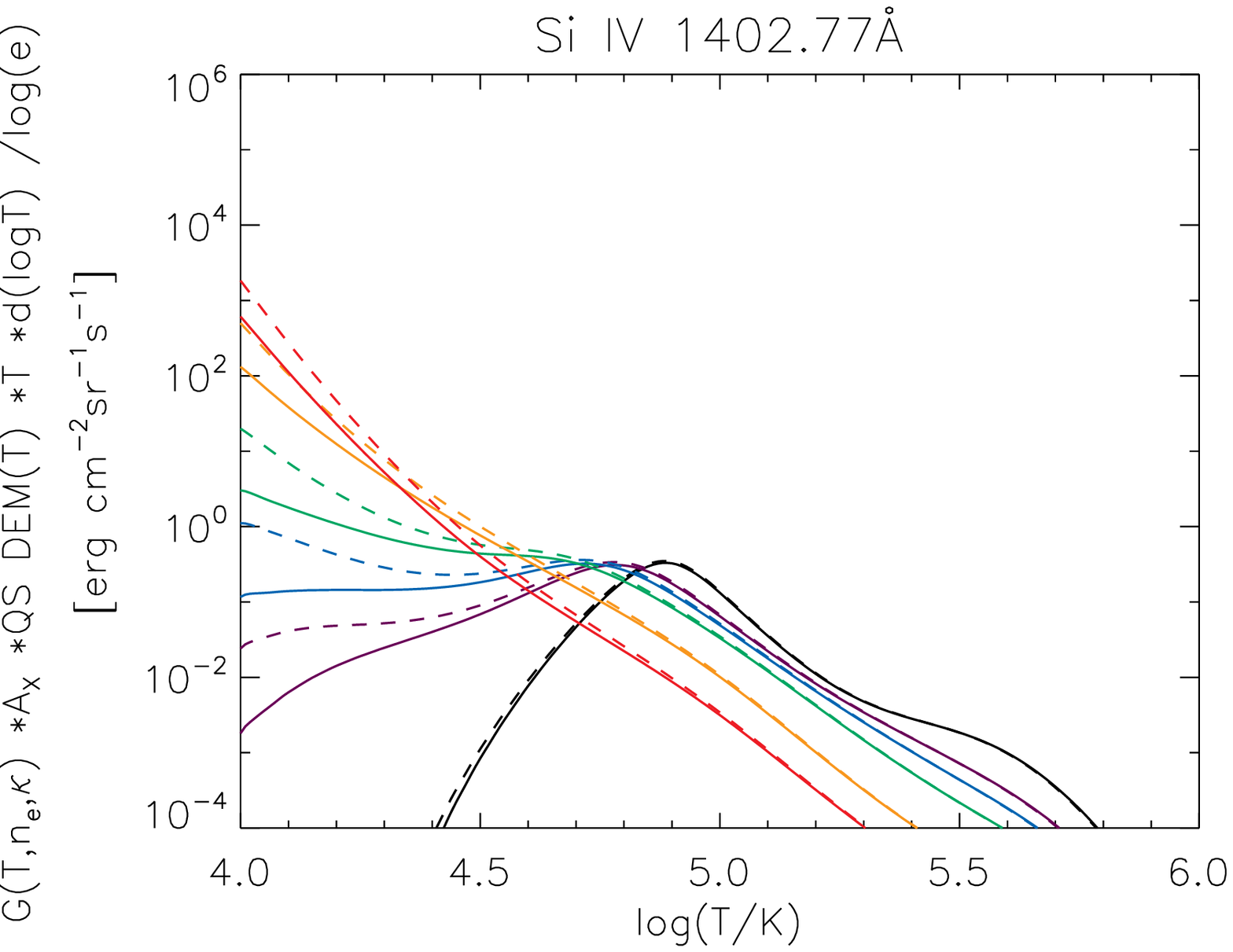}
       \includegraphics[width=7.8cm,clip]{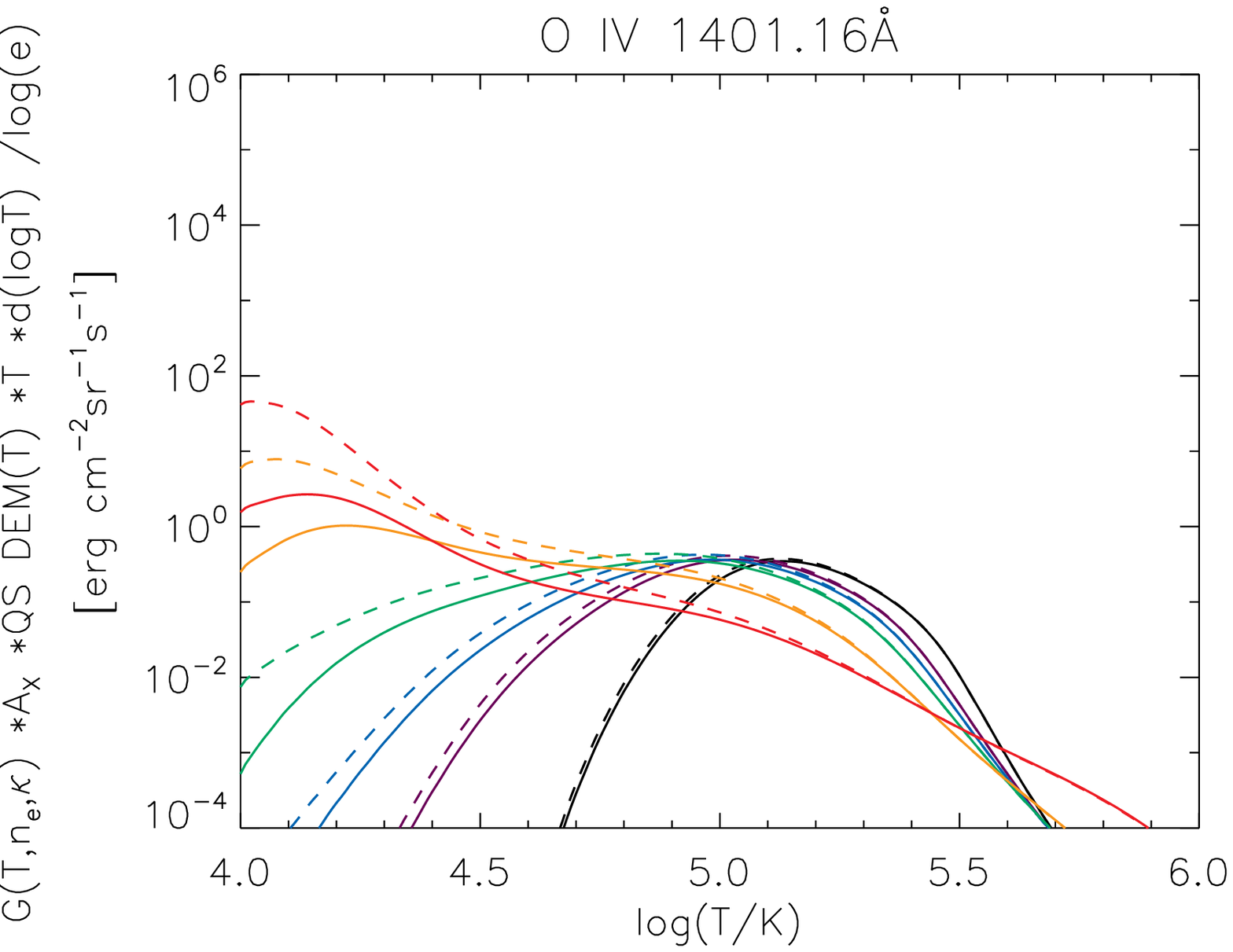}
       \includegraphics[width=7.8cm,clip]{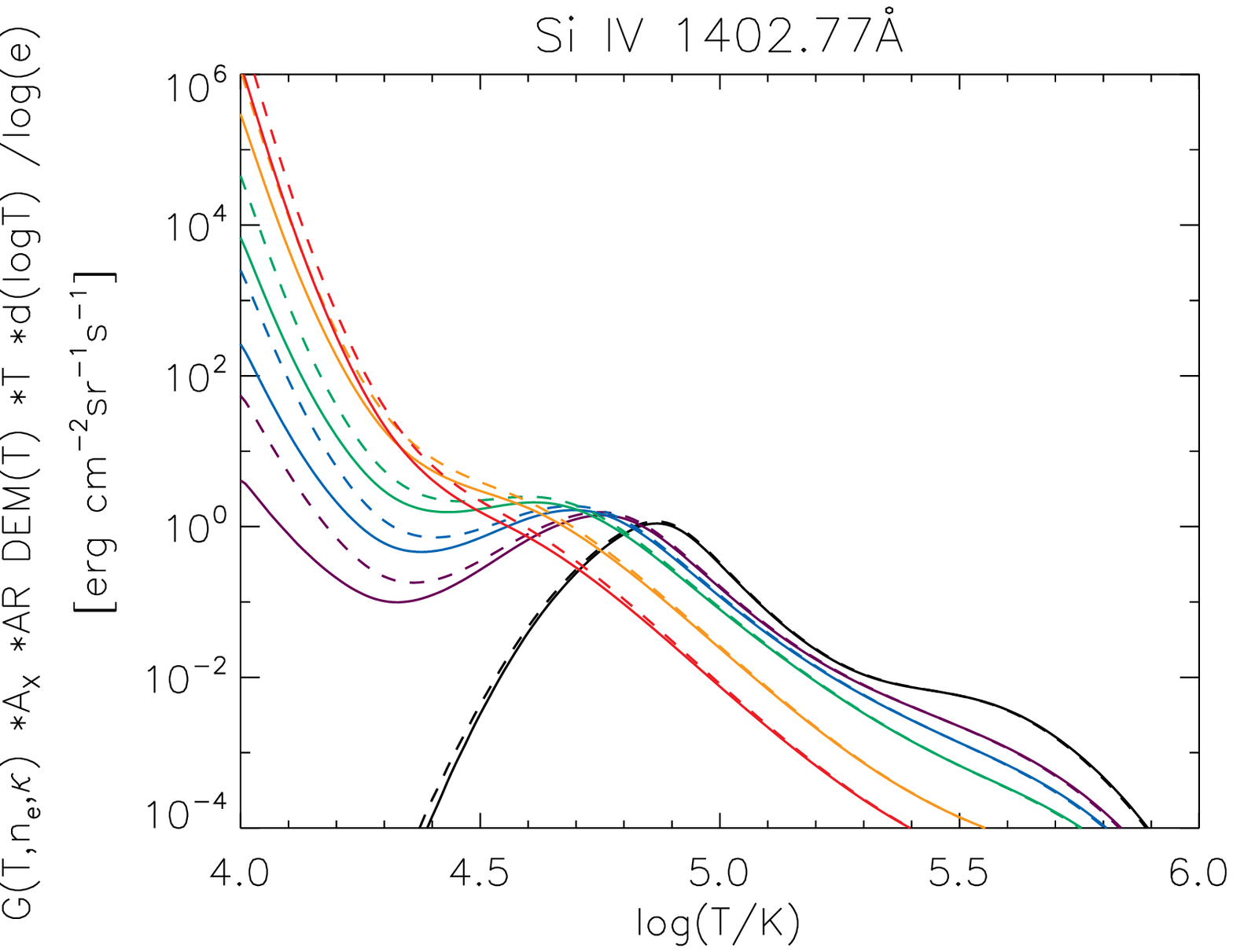}
       \includegraphics[width=7.8cm,clip]{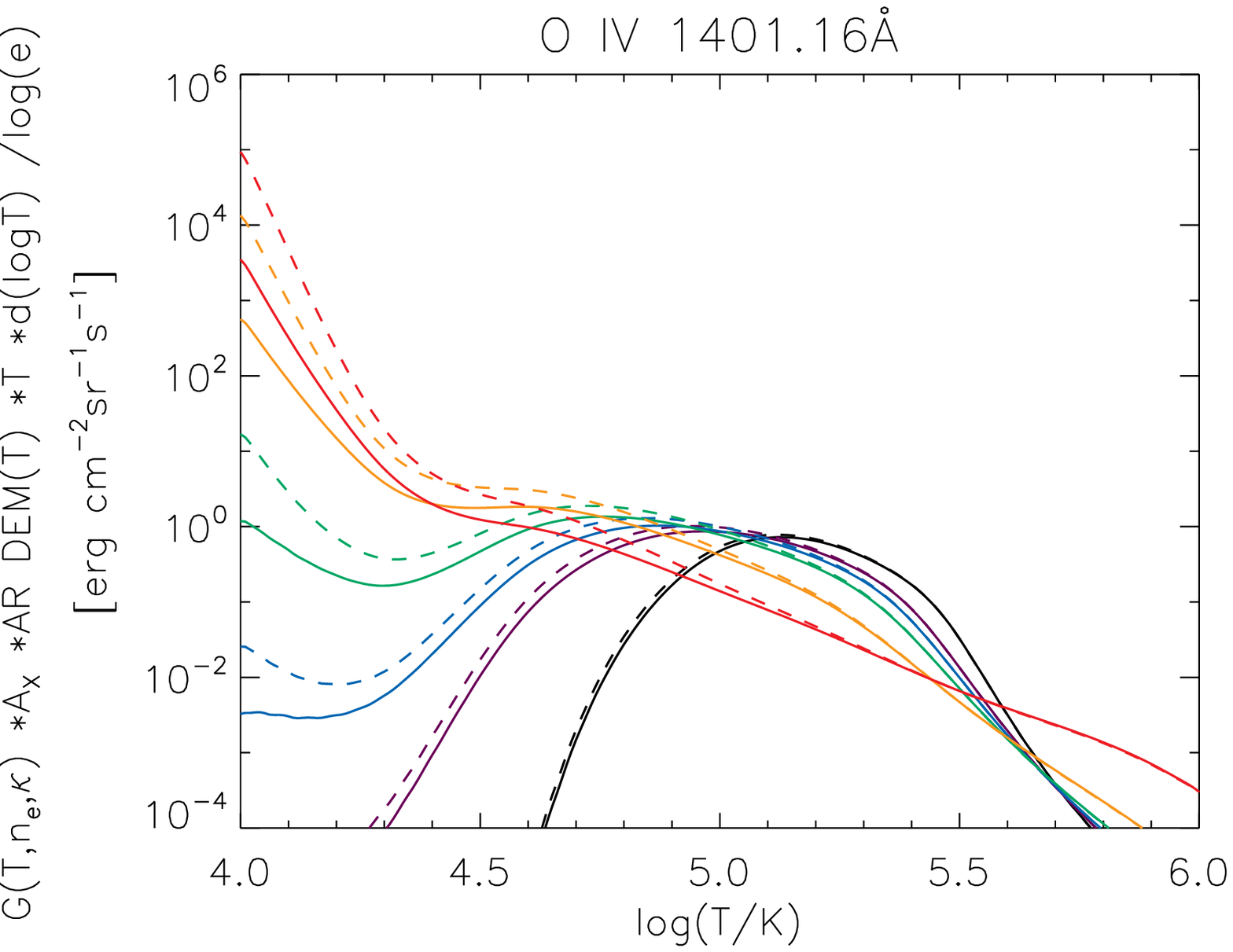}
     \caption{Contribution functions $G(T,n_\mathrm{e},\kappa)$ for the \ion{Si}{4} 1402.77\AA~(\textit{Left}) and \ion{O}{4} 1401.16\AA~transitions (\textit{Right}). The \textit{middle} and \textit{bottom} rows depict the contribution functions multiplied by the d$T$ factors and the quiet-Sun and active-region DEMs, respectively. (A color version of this figure is available in the online journal.)}
       \label{Fig:contrib_funct}
   \end{figure*}

The $G_{ji}(T,n_\mathrm{e},\kappa)$ of the \ion{O}{4} 1401.16\AA~and \ion{Si}{4} 1402.77\AA~are shown in Fig. \ref{Fig:contrib_funct}, \textit{top}. We chose these two lines since they are close in wavelength and belong to the strongest TR lines observed by IRIS. The contribution functions are calculated either under the assumption of constant pressure $P_\mathrm{e}$\,=\,3$\times$10$^{15}$\,Kcm$^{-3}$, or constant density of log$(n_\mathrm{e}/\mathrm{cm}^{-3})$\,=\,10. The first assumption is appropriate for a locally ``open'' magnetic structure, such as a TR portion of a coronal loop footpoint, while the second one is neccessary for investigating the density sensitivity (Sect. \ref{Sect:3.3}). Note that the same value of $P_\mathrm{e}$ was used to derive the DEM files within CHIANTI \citep{Dere97}.

With decreasing $\kappa$, i.e., increasing departure from the Maxwellian distribution, the contribution functions for both lines becomes broader in temperature and their maxima move towards lower $T$. This is caused chiefly by the changes in the  relative ion abundances (Fig. \ref{Fig:atomic_data}, \textit{top}). However, temperatures corresponding to the $G(T,n_\mathrm{e},\kappa)$ peak are not as low as the temperatures at which the \ion{O}{4} and \ion{Si}{4} ions have maximum abundance. This is caused by the level population, which increases with $T$ in the temperature interval where the ion is formed. For log$(n_\mathrm{e}/\mathrm{cm}^{-3})$\,=\,10, the \ion{O}{4} 1401.16\AA\,line has peak formation temperature at log$(T/\mathrm{K})$\,$\approx$\,5.15 for the Maxwellian distribution, and at $\approx$\,4.8 for $\kappa$\,=\,2. For \ion{Si}{4}, these values are $\approx$\,4.9 and 4.3, respectively. We note that the peak formation temperature for the Maxwellian distribution is significantly higher than the one given by \citet[][Fig. 1 therein]{Doyle13}. The peak temperatures obtained here for both the \ion{Si}{4} and \ion{O}{4} are largely independent of $n_\mathrm{e}$. However, the contribution function does change with $n_\mathrm{e}$ (Sect. \ref{Sect:3.3}). For this reason, the peak temperatures are located at slightly higher $T$ if constant pressure is assumed (Fig. \ref{Fig:contrib_funct}, \textit{top}).

We investigated the \ion{O}{4} line ratios for possible diagnostics of $\kappa$. Since the lines are formed from similar energy levels (Table \ref{Table:1}), ratios of these lines do not allow for simultaneous diagnostics of $T$ and $\kappa$. Lines with wavelengths shorter or longer by several tens or hundreds of \AA~would be required for direct diagnostics of $\kappa$ \citep[c.f.][]{Dzifcakova11Si,Mackovjak13}. In particular, combination of the IRIS \ion{O}{4} lines with the ones observed by \textit{SOHO}/SUMER can lead to diagnostics of $\kappa$. We recommend using e.g. the 554.51\AA, 787.71\AA, and 790.20\AA~lines. As an example, the 1401.16\AA\,/\,787.71\AA~ratio in combination with the 790.20\AA\,/\,1399.77\AA~ratio could lead to simultaneous diagnostics of $T$ and $\kappa$, if $n_\mathrm{e}$ is known. Note that combinations with the \textit{Hinode}/EIS \ion{O}{4} lines (e.g., 279.93\AA) could lead to very sensitive diagnostics, but these lines are weak even in long exposures \citep{Brown08}. However, calibration and cross-calibration uncertainties limit the usefulness of the line ratio method, leaving modelling the entire observed \ion{O}{4} spectrum \citep[as done for \ion{Si}{3} in][]{Dzifcakova11Si} as the most viable option for determining $\kappa$.

%
%
\subsection{Effect of DEM on line formation}
\label{Sect:3.2}

The shifts of the peak temperatures of the contribution function with $\kappa$ is potentially important, given the slope of the DEM at TR temperatures. To illustrate the effect of DEM on the contribution function, we chose the ``Active region'' (AR) and ``Quiet Sun'' (QS) DEMs available within CHIANTI v7.1, based on the data of \citet{Vernazza78a}. The AR DEM has a much steeper transition region than the QS DEM. We note that the spectra used to produce these DEMs are averages over many exposures and thus represent average physical conditions in the solar atmosphere over a selected type of structure on the solar surface; e.g., a quiet-Sun or an active region. This means that such DEMs may not be representative of any \textit{particular}, sub-arcsecond feature observed by IRIS. Moreover, the DEMs were derived under the assumption of a Maxwellian distribution, so that using them in conjunction with finite $\kappa$ is not self-consistent. Nevertheless, we use these DEMs to illustrate the general effect of the DEM slope on the formation of the IRIS TR lines. We note that the true DEMs for $\kappa$-distributions may differ mainly for very low temperatures due to the behaviour of $G(T,n_\mathrm{e},\kappa)$ with $\kappa$ described in Sect. \ref{Sect:3.1}. However, evaluating the DEMs for $\kappa$-distributions is beyond the scope of this paper.

The results are shown in Fig. \ref{Fig:contrib_funct} \textit{middle} and \textit{bottom}. In this figure, the contribution functions are also multiplied by the d$T$ = $T \mathrm{d}(\mathrm{log}T)/\mathrm{log(e)}$ factor, so that the intervals of log$(T/$K) which contribute most to the total intensity can be immediately discerned. It is seen that for $\kappa$\,=\,2--7 and the QS DEM, the \ion{Si}{4} 1402.77\AA\,line is dominantly formed in regions with very low temperature: log$(T/\mathrm{K}) \to$\,4. The situation is even more pronounced for the AR DEM, where the line is formed at such low temperature even for $\kappa$\,=\,10. The situation is different for the \ion{O}{4} 1401.16\AA\,line, which is formed at very low $T$ only for $\kappa$\,$\lesssim$\,3 for QS DEM and $\kappa$\,$\lesssim$\,5 for AR DEM. For higher $\kappa$, the contribution function is slightly shifted to lower $T$ and has a pronounced low-temperature wing.

\begin{figure}
       \centering
       \includegraphics[width=8.8cm,clip]{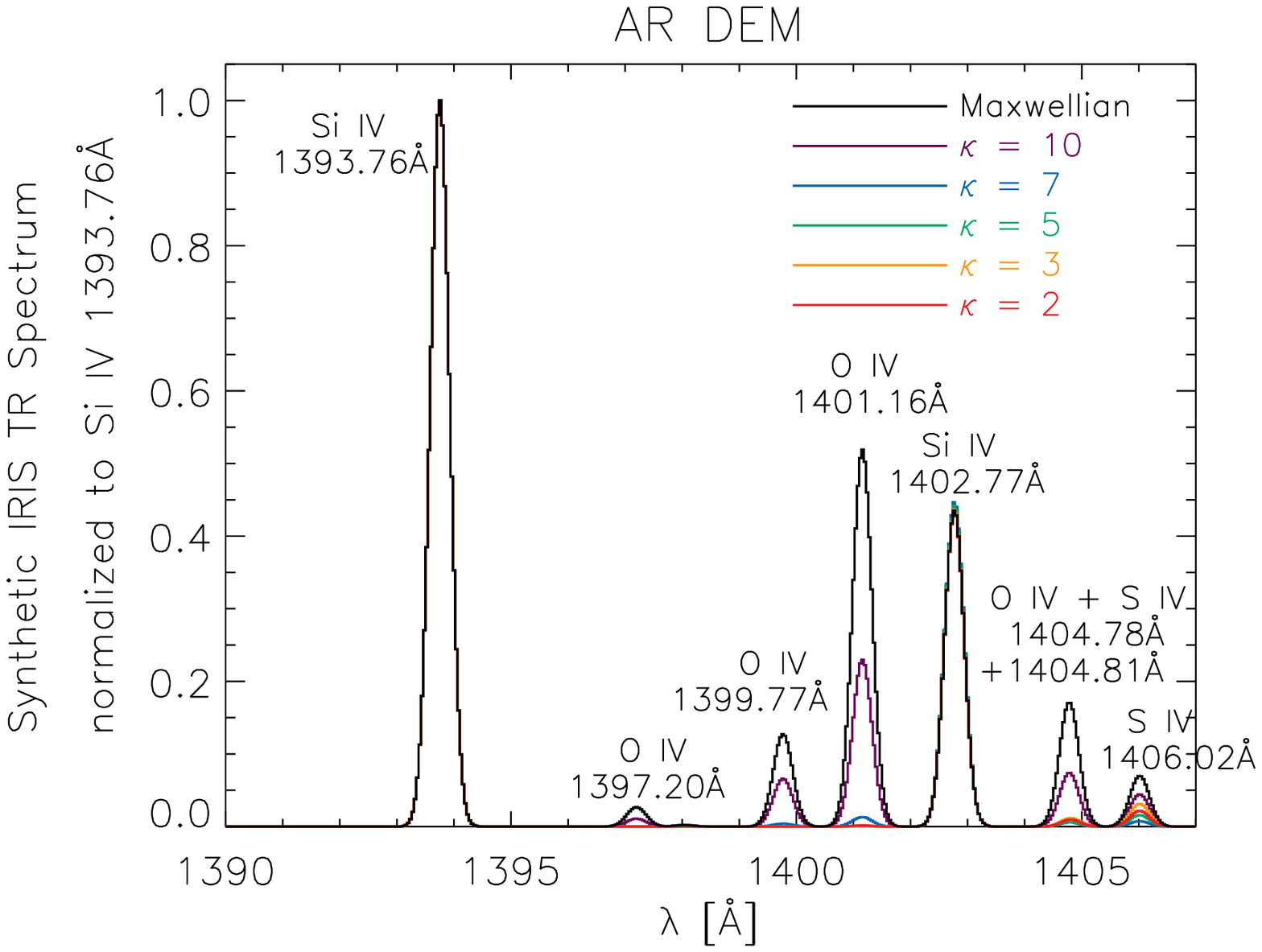}
       \includegraphics[width=8.8cm,clip]{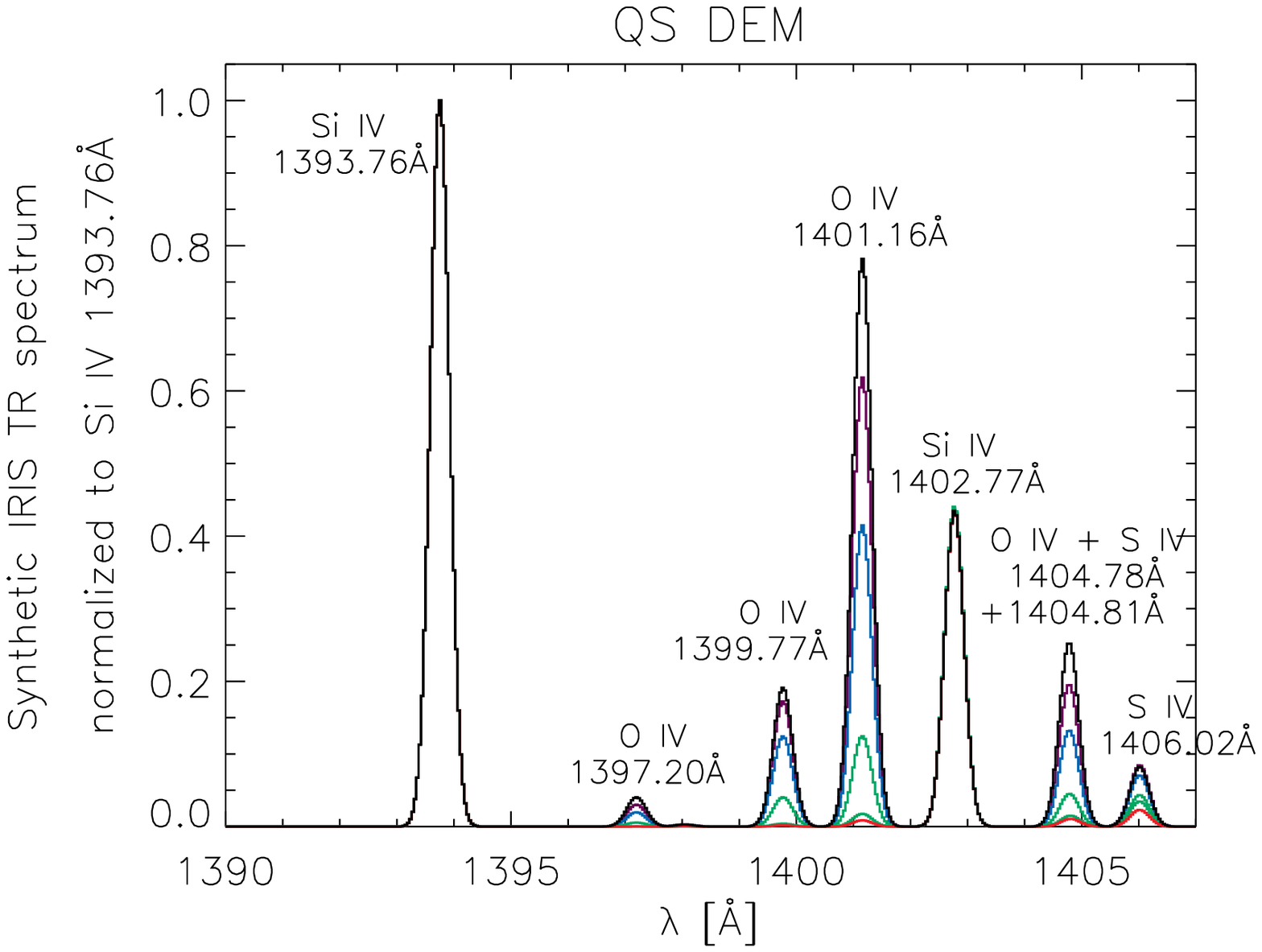}
     \caption{Sythetic TR line spectra of \ion{O}{4}, \ion{Si}{4} and \ion{S}{4} in the second FUV channel of IRIS, calculated for the AR or QS DEMs.(A color version of this figure is available in the online journal.)}
       \label{Fig:DEM_spectra}
   \end{figure}

The synthetic spectra arising in regions characterized by such DEMs are shown in Fig. \ref{Fig:DEM_spectra}. These spectra have been scaled by the IRIS effective area and the FWHM of IRIS lines have been artificially increased so that the line intensities for different $\kappa$ are better visible. We note here that the \ion{Si}{4} 1393.76\AA\,/1402.77\AA~ratio is approximately constant and equal to 2 independently of $\kappa$ or any other assumptions. This is in agreement with the SUMER results presented by \citet{Doschek01}. As shown in Fig. \ref{Fig:DEM_spectra}, the \ion{O}{4} lines are strongly sensitive to $\kappa$. For the AR DEM, they are very weak except for $\kappa$\,=\,10. For the QS DEM, they are weaker compared to Maxwellian distribution, but still observable even for $\kappa$\,=\,5.

We note that given the possible values of $\kappa$, which are $\approx$\,7 for the active region and $\approx$\,9--12 for quiet-Sun \citep{Dzifcakova11Si}, the \ion{Si}{4} and \ion{O}{4} lines may be formed under very different conditions, i.e., in different structures. These differences might be resolvable by IRIS. The effect of finite $\kappa$ would be to move the \ion{Si}{4} emission into the low TR closer to the chromosphere, i.e., into lower-lying, more dense structures that are possibly more narrow due to constriction by the expanding magnetic flux-tubes. In this scenario, the \ion{O}{4} emission would lie higher up, in the region where log$(T/$K)\,$\approx$\,5. We finally note that the relative intensities might be modified by abundance effects, especially for \ion{Si}{4}, which is a low-FIP element.

%
%
\subsection{Electron density diagnostics}
\label{Sect:3.3}

\begin{figure*}[!ht]
       \centering
       \includegraphics[width=7.8cm,clip]{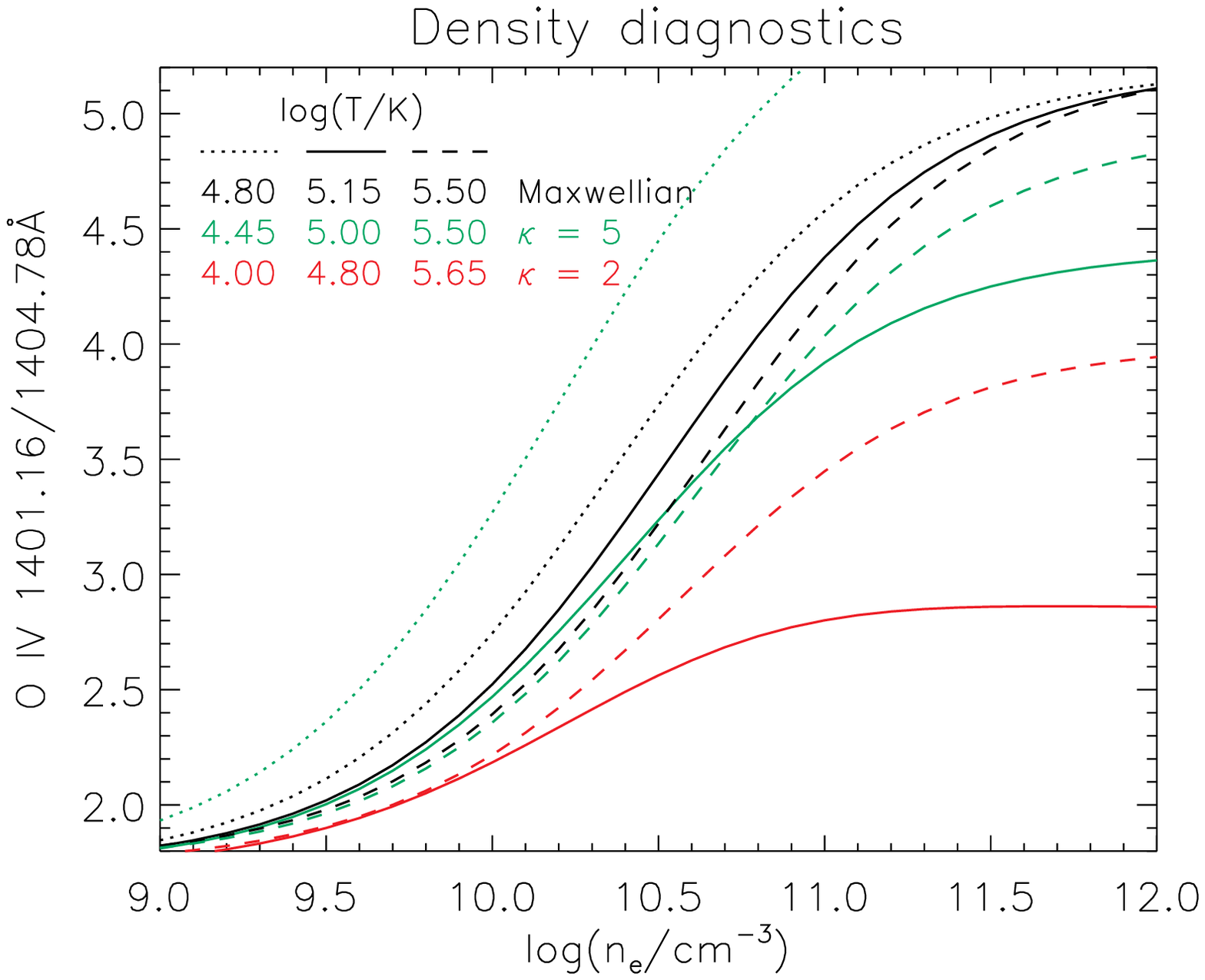}
       \includegraphics[width=7.8cm,clip]{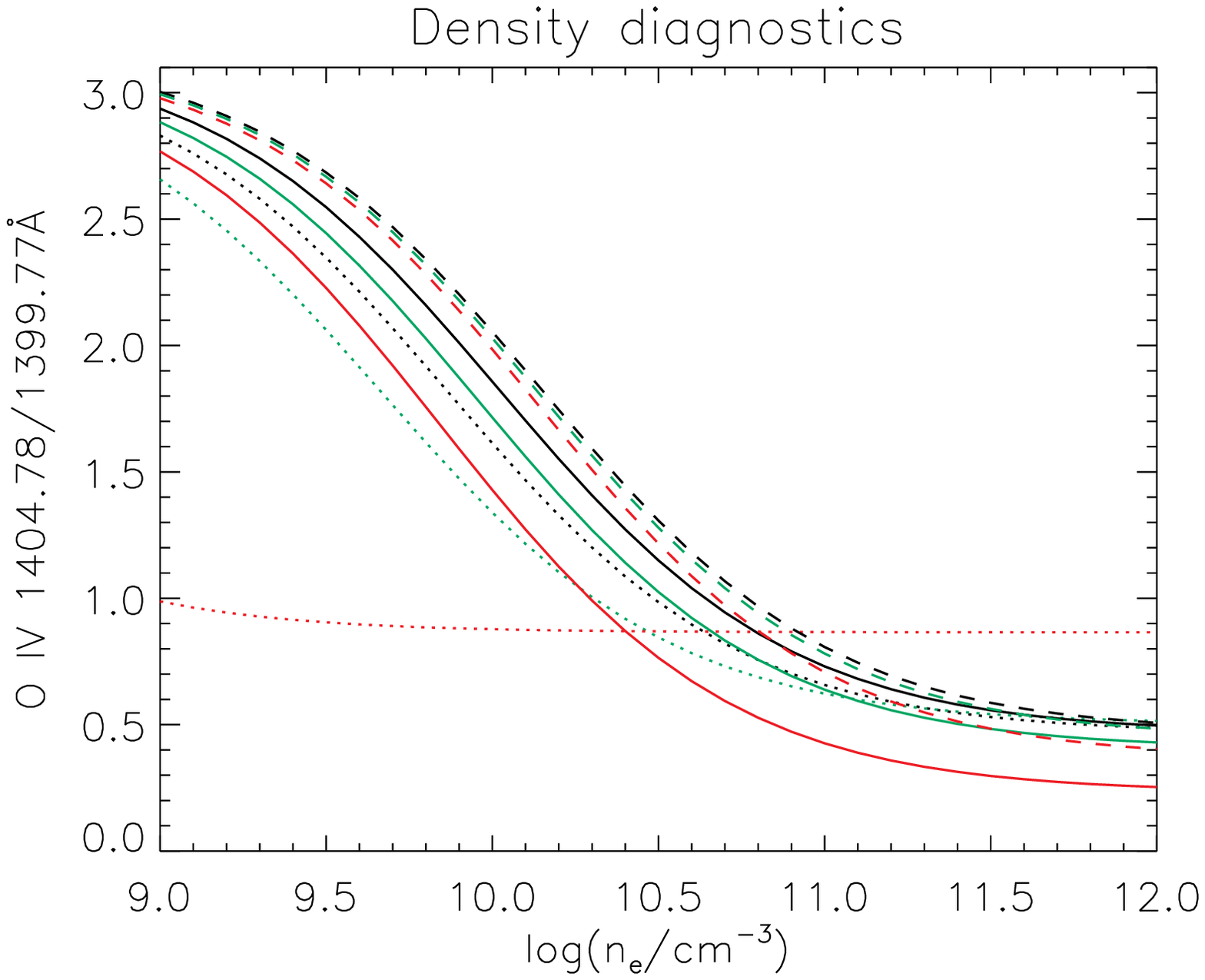}
       \includegraphics[width=7.8cm,clip]{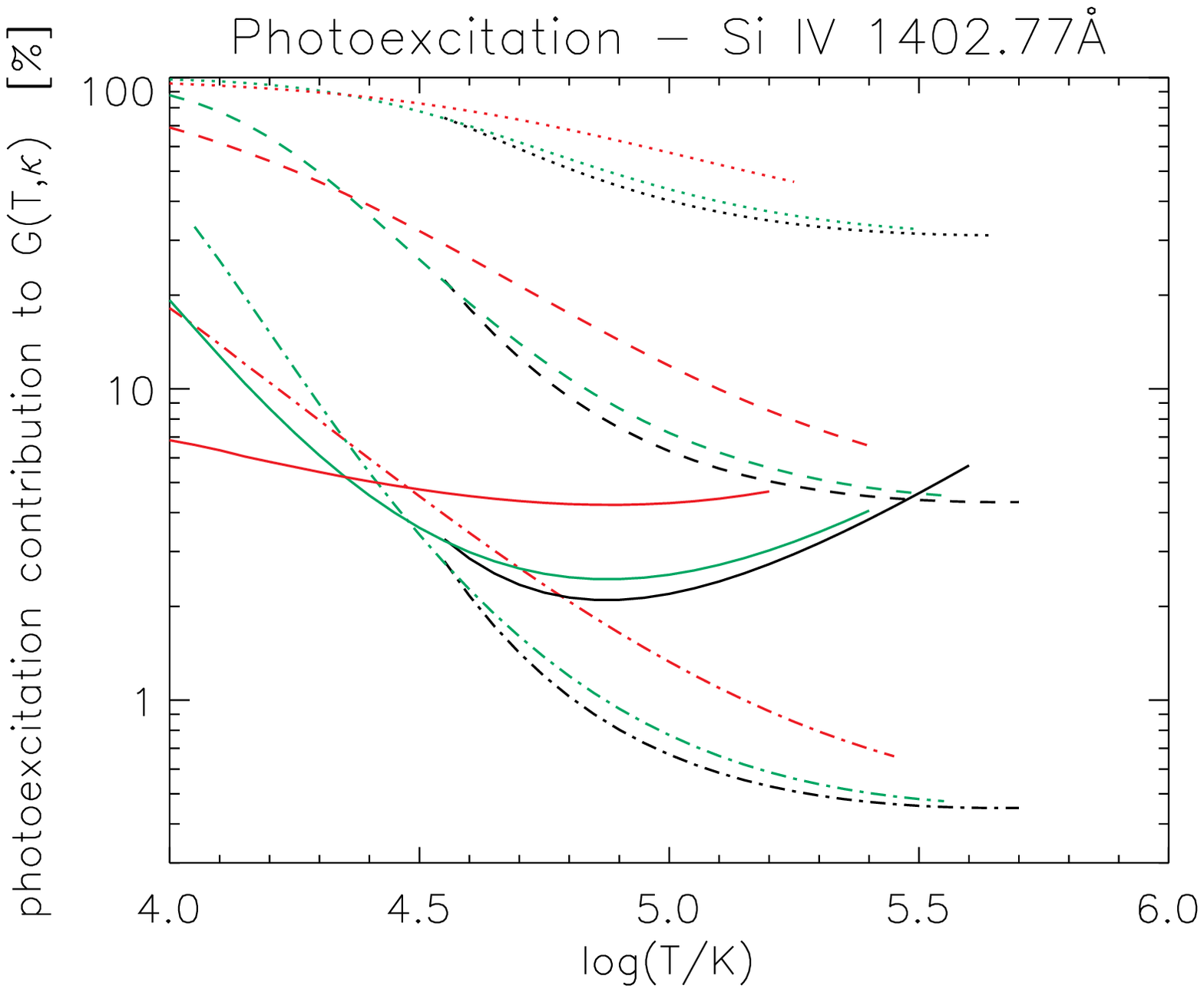}
       \includegraphics[width=7.8cm,clip]{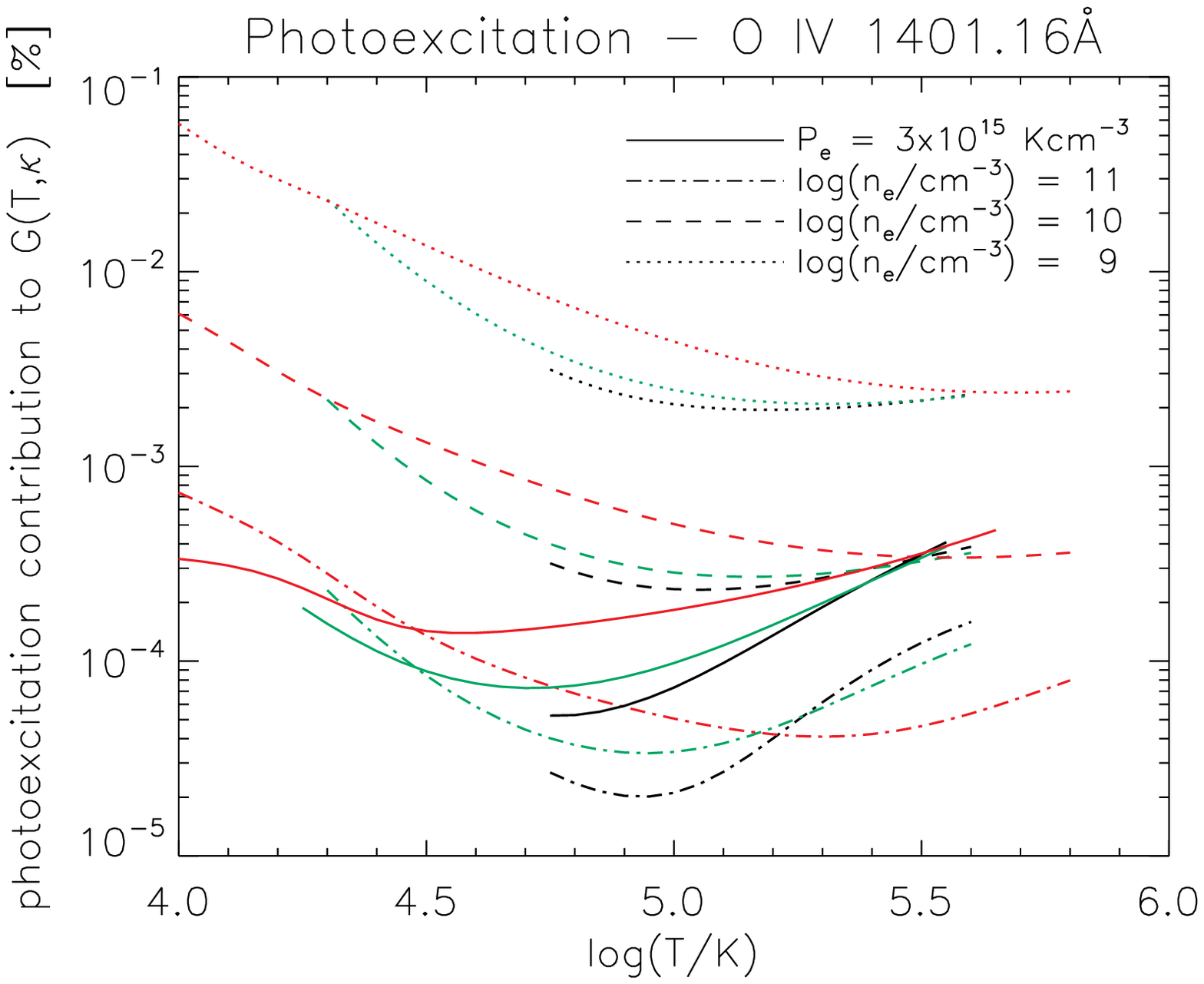}
     \caption{\textit{Top}: Density diagnostics using \ion{O}{4} line ratios. Individual line styles correspond to different log$(T/$K). Colors correspond to different $\kappa$. \textit{Bottom}: Relative contribution (in \%) from photoexcitation to the total $G(T,n_\mathrm{e},\kappa)$ for the \ion{Si}{4} 1402.77\AA~(\textit{left bottom}) and \ion{O}{4} 1401.16\AA\,lines (\textit{right bottom}). (A color version of this figure is available in the online journal.)}
       \label{Fig:photoexcit}
   \end{figure*}

It is known that the intercombination multiplet of \ion{O}{4} can be used for diagnostics of $n_\mathrm{e}$ \citep[e.g.,][]{Vernazza78b,Feldman81,Nussbaumer82,Doschek84,Hayes87,Dwivedi92,Keenan09}. Here, we investigate the effect of $\kappa$ on the density diagnostics. To do that, we plot the $n_\mathrm{e}-$dependence of a chosen ratio of \ion{O}{4} lines observable by IRIS (Table \ref{Table:1}) for the peak formation temperature and temperatures corresponding to the 1\% of the contribution function peak. This gives the indication of the sensitivity of the given ratio to $T$ (Fig. \ref{Fig:photoexcit}, \textit{top}).

If the Maxwellian distribution is assumed, the \ion{O}{4} line ratios yield a good density diagnostics in the log$(n_\mathrm{e}/\mathrm{cm}^{-3})$\,=\,9$-$11 range (black lines in Fig. \ref{Fig:photoexcit}, \textit{top}), with lines at different $T$ producing variations of about $\pm$0.2 in log$(n_\mathrm{e}/\mathrm{cm}^{-3})$.  At even higher densities, the ratios reach a plateau and are unusable for log$(n_\mathrm{e}/\mathrm{cm}^{-3})$\,$>$\,11.5.

In principle, the density diagnostics can depend on $\kappa$, but the exact manner can be different for each ratio. The 1401.16\AA\,/1404.78\AA~ratio is not recommended for density diagnostics, since it changes with $\kappa$ (Fig. \ref{Fig:photoexcit}, \textit{top left}). On the other hand, the 1404.78\AA\,/1399.77\AA~ratio is only weakly dependent on $\kappa$ for $\kappa$\,$\gtrsim$\,5, yielding slightly lower $n_\mathrm{e}$. For the extreme non-Maxwellian situation, i.e., $\kappa$\,=\,2, the ratio is useful only for temperatures close to or higher the peak formation temperature. For very low log$(T$/K)\,$\to$\,4, the density dependence is lost (Fig. \ref{Fig:photoexcit}, \textit{top right}).

We note that the \ion{O}{4} 1404.78\AA~line is blended with the \ion{S}{4} 1404.81\AA~transition. This blend can become important at higher densities, since the intensity of the \ion{O}{4} 1404.78\AA~line decreases with $n_\mathrm{e}$. The contribution of this blend can in principle be estimated using the neighboring \ion{S}{4} 1406.02\AA~line, since the ratio of these two \ion{S}{4} lines is insensitive to $n_\mathrm{e}$ except for very high densities of log$(n_\mathrm{e}/\mathrm{cm}^{-3})$\,$\gtrsim$\,12 \citep[][Fig. 1 therein]{Doschek84}. We note that the peak formation temperature of these \ion{S}{4} is log$(T/$K)\,=\,5.0 for the Maxwellian distribution and 4.6 for $\kappa$\,=\,2.

%
\subsection{Photoexcitation}
\label{Sect:3.4}

To quantify the effect of photoexcitation, we recalculated the contribution functions without it. It is found that the \ion{O}{4} lines are unaffected by photoexcitation, which contributes less than 0.1\% to the \ion{O}{4} 1401.16\AA~line (Fig. \ref{Fig:photoexcit}, \textit{bottom}).

The situation is different for the \ion{Si}{4} 1402.77\AA\,line. Here, the effect of photoexcitation is dependent on the conditions under which the line is formed. For constant $P_\mathrm{e}$\,=\,3\,$\times$10$^{15}$\,K\,cm$^{-3}$, photoexcitation contributes 2--7\%. For constant density, the relative contribution depends on $T$, $n_\mathrm{e}$, and $\kappa$. In general, the relative contribution increases with decreasing $T$, $\kappa$, or $n_\mathrm{e}$ (Fig. \ref{Fig:photoexcit}). It can be larger than 10\% for log$(n_\mathrm{e}/\mathrm{cm}^{-3})$\,=\,10, and much larger still for even lower $n_\mathrm{e}$. We conclude that the effect of photoexcitation cannot be neglected for the \ion{Si}{4} lines, and suggest that a real radiation field should be used instead of a black-body spectrum as done here.

%
%
%
\section{Summary}
\label{Sect:4}

We have investigated the formation of the \ion{O}{4}, \ion{Si}{4}, and \ion{S}{4} lines observable in the second far-ultraviolet channel of the Interface Region Imaging Spectrograph (IRIS). To do that, we utilized the state-of-art atomic data that will be part of the upcoming CHIANTI version 8. The \ion{O}{4} 1401.16\AA~line is formed at higher temperatures than the neighboring \ion{Si}{4} 1402.77\AA\,line. This is true for all values of $\kappa$ considered. Taking into account the slope of the DEM in the transition region, the \ion{Si}{4} line can be formed predominantly at very low temperatures of log$(T/$K)\,$\to$\,4 even for weakly non-Maxwellian situations. In contrast, the \ion{O}{4} 1401.16\AA\,line will be formed at such low $T$ only under extremely non-Maxwellian situations of $\kappa$\,$\approx$\,2--3 or in regions characterized with very steep DEM slopes. If the values of $\kappa$ diagnosed by \citet{Dzifcakova11Si} are correct, the \ion{Si}{4} lines will be formed in different parts of the transition region than the \ion{O}{4} line, a situation possibly discernible by the IRIS instrument. The synthetic spectra are predicted to have lower \ion{O}{4} intensities compared to the purely Maxwellian transition region. No direct diagnostics of $\kappa$ was found using \ion{O}{4} line ratios, due to close wavelengths of these lines, which are formed from upper levels having similar energies.

The usefulness of the \ion{O}{4} lines for density diagnostics depends on the ratio considered. The \ion{O}{4} 1404.78\AA\,/1399.77\AA~ratio is particularly useful, as it is only weakly dependent on temperature and $\kappa$, except for extremely non-Maxwellian situation of $\kappa$\,$\approx$\,2 and very low log$(T/$K)\,$\to$\,4, where the density sensitivity is lost. However, the \ion{O}{4} 1404.78\AA\,line is blended with the \ion{S}{4} 1404.81\AA\,transition. The contribution of this blend can be estimated using the \ion{S}{4} 1406.02\AA\,line.

It is also found that photoexcitation of the \ion{Si}{4} lines by photospheric and/or chromospheric radiation cannot be neglected. We recommend using realistic spectra to calculate the photoexcitation contribution to these lines. The \ion{O}{4} lines are not found to be sensitive to photoexcitation.

\acknowledgments

The authors thank Martin O'Mullane for handling the APAP data. JD acknowledges support from the Royal Society via the Newton Fellowship Programme, and from Scientific Grant Agency (VEGA), Slovakia, Grant No. 1/0240/11. GDZ and HEM acknowledge support from STFC. ED acknowledges the Grant Agency of the Czech Republic, Project No. P/209/12/1652. CHIANTI is a collaborative project involving the NRL (USA), the University of Cambridge (UK), and George Mason University (USA). Scientific and OpenSuse Linux operating systems were used to produce this work.



\clearpage
%
\begin{table}[!h]
\begin{center}
\caption{Transition-region lines in the second far-ultraviolet channel of the IRIS spacecraft (1390-1406\AA).
\label{Table:1}}
\begin{tabular}{cccc}
\tableline\tableline
ion		& $\lambda$ [\AA]& \hspace{0.5cm} transition					& \hspace{0.3cm} levels	 \\
\tableline\tableline
\ion{Si}{4}	& 1393.76 & \hspace{0.5cm} $3s~~ {}^2 S_{1/2} - 3p~~ {}^2 P_{3/2}$	 & \hspace{0.3cm} $1 - 3$ \\
\ion{Si}{4}	& 1402.77 & \hspace{0.5cm} $3s~~ {}^2 S_{1/2} - 3p~~ {}^2 P_{1/2}$	 & \hspace{0.3cm} $1 - 2$ \\
\tableline
\ion{O}{4}	& 1397.20 & \hspace{0.5cm} $2s^2 2p~~ {}^2P_{1/2} - 2s 2p^2~~ {}^4 P_{3/2}$ & \hspace{0.3cm} $1 - 4$ \\
\ion{O}{4}	& 1399.77 & \hspace{0.5cm} $2s^2 2p~~ {}^2P_{1/2} - 2s 2p^2~~ {}^4 P_{1/2}$ & \hspace{0.3cm} $1 - 3$ \\
\ion{O}{4}	& 1401.16 & \hspace{0.5cm} $2s^2 2p~~ {}^2P_{3/2} - 2s 2p^2~~ {}^4 P_{5/2}$ & \hspace{0.3cm} $2 - 5$ \\
\ion{O}{4}	& 1404.78 & \hspace{0.5cm} $2s^2 2p~~ {}^2P_{3/2} - 2s 2p^2~~ {}^4 P_{3/2}$ & \hspace{0.3cm} $2 - 4$ \\
\tableline
\ion{S}{4}	& 1398.04 & \hspace{0.5cm} $3s^2 3p~~ {}^2 P_{1/2} - 3s 3p^2~~ {}^4 P_{3/2}$ & \hspace{0.3cm} $1 - 4$ \\
\ion{S}{4}	& 1404.81 & \hspace{0.5cm} $3s^2 3p~~ {}^2 P_{1/2} - 3s 3p^2~~ {}^4 P_{1/2}$ & \hspace{0.3cm} $1 - 3$ \\
\ion{S}{4}	& 1406.02 & \hspace{0.5cm} $3s^2 3p~~ {}^2 P_{3/2} - 3s 3p^2~~ {}^4 P_{5/2}$ & \hspace{0.3cm} $2 - 5$ \\
\tableline\tableline
\end{tabular}
\end{center}
\end{table}

\clearpage
%

\clearpage
%

\clearpage
%

\clearpage
%

\end{document}